\documentclass[aps,prd,tightenlines,preprintnumbers,superscriptaddress,twocolumn,notitlepage,nofootinbib,eqsecnum,floatfix]{revtex4-1}

\usepackage{graphicx}
\usepackage{amsmath}
\usepackage[pdftex]{color}
\usepackage{hyperref}
\hypersetup{
    colorlinks=true,       
    linkcolor=blue,          
    citecolor=blue,        
    filecolor=blue,      
    urlcolor=blue           
}

\newcommand{\order}{\ensuremath{\text{O}}} 
\newcommand{\Pihat}{\widehat\Pi}
\newcommand{\gmtwo}{$g_\mu-2$}
\newcommand{\amu}{$a_\mu^{\rm HVP}$}
\newcommand{\Pade}{Pad{\'e}}
\newcommand{\be}{\begin{equation}}
\newcommand{\ee}{\end{equation}}
\newcommand{\alphaEM}{\alpha_{\rm EM}}
\newcommand{\bea}{\begin{eqnarray}}
\newcommand{\eea}{\end{eqnarray}}

\newcommand{\st}[1]{{}}                                 
\definecolor{grey}{rgb}{0.6,0.6,0.6}

\begin{document}
\preprint{FERMILAB-PUB-18-287-T}

\widetext

\title{Higher-order hadronic-vacuum-polarization contribution \\ to the muon $g-2$ from lattice QCD}
\author{B.~Chakraborty}
\affiliation{Jefferson Lab, 12000 Jefferson Avenue, Newport News, Virginia 23606, USA}

\author{C.~T.~H.~Davies}
\affiliation{SUPA, School of Physics and Astronomy, University of Glasgow, Glasgow, G12 8QQ, UK}

\author{J.~Koponen} 
\affiliation{INFN, Sezione di Roma Tor Vergata, Via della Ricerca Scientifica 1, 00133 Roma RM, Italy}

\author{G.~P.~Lepage} 
\affiliation{Laboratory for Elementary-Particle Physics, Cornell University, Ithaca, New York 14853, USA}

\author{R.~S.~\surname{Van de Water}}\email{ruthv@fnal.gov}
\affiliation{Fermi National Accelerator Laboratory, Batavia, Illinois, 60510, USA}

\collaboration{Fermilab Lattice, HPQCD, and MILC Collaborations}
\noaffiliation

\date{\today}

\begin{abstract}
We introduce a new method for calculating the $\order(\alpha^3)$ hadronic-vacuum-polarization contribution to the muon anomalous magnetic moment from {\it ab-initio} lattice QCD. 
We first derive expressions suitable for computing the higher-order contributions either from the renormalized vacuum polarization function $\Pihat(q^2)$, or directly from the lattice vector-current correlator in Euclidean space.
We then demonstrate the approach using previously-published results for the Taylor coefficients of $\Pihat(q^2)$  that were obtained on four-flavor QCD gauge-field configurations with physical light-quark masses.  
We obtain $10^{10} a_\mu^{\rm HVP,HO} = -9.3(1.3)$, in agreement with, but with a larger uncertainty than, determinations from $e^+e^- \to {\rm hadrons}$ data plus dispersion relations. 
\end{abstract}

\maketitle

\section{Introduction}
\label{sec:intro}

The anomalous magnetic moment of the muon (\gmtwo) is one of the most precisely-determined observables in particle physics, having been measured with an uncertainty of 0.54 parts-per-million by BNL Experiment E821~\cite{Bennett:2006fi}.   
Because of this high experimental precision,  and because the anomaly is mediated by quantum-mechanical loops in the Standard Model, the muon \gmtwo\ provides stringent constraints on new heavy or weakly-coupled particles.  
The present Standard-Model theory value lies below the BNL E821 measurement by more than three standard deviations~\cite{Blum:2013xva}.  
To identify definitively whether this deviation is due to new particles or forces, both the theory and measurement errors must be improved.
The Muon \gmtwo\ Experiment recently began running at Fermilab, and aims to reduce experimental error by a factor of four~\cite{Grange:2015fou}.
In parallel, numerous efforts are underway by the lattice-QCD community to tackle the Standard-Model hadronic contributions~\cite{Aubin:2006xv,Boyle:2011hu,Burger:2013jya,Blum:2015you,Chakraborty:2016mwy,Blum:2016lnc,Borsanyi:2016lpl,DellaMorte:2017dyu,Borsanyi:2017zdw,Blum:2018mom}, which are the largest source of theory uncertainty~\cite{Blum:2013xva}.  

The largest source of uncertainty in the Standard-Model \gmtwo\ is from the \order($\alpha^2$) hadronic vacuum-polarization (HVP) contribution~\cite{Blum:2013xva}, \amu, which is shown in Fig.~\ref{fig:HVP}.\footnote{The symbol $\alpha$ always denotes the electromagnetic coupling in this work.}  
This contribution can be obtained by combining experimental measurements of electron-positron inclusive scattering into hadrons with dispersion relations, and recent determinations from this approach quote errors of  0.4--0.6\%~\cite{Jegerlehner:2017lbd,Davier:2017zfy,Keshavarzi:2018mgv}.  
The most precise calculation of the leading-order \amu\ to-date from Ref.~\cite{Chakraborty:2016mwy} employed four-flavor lattice QCD with physical-mass pions to achieve a total error of $\sim$ 2\%.  A significant source of systematic uncertainty in this and all lattice-QCD results to-date is from the use of degenerate up- and down-quark masses; phenomenological estimates of this error are about~1\%~\cite{Wolfe:2010gf,Jegerlehner:2011ti,Jegerlehner:2017gek}.  Recently, we calculated the strong-isospin-breaking correction to the leading-order, light-quark-connected contribution to \amu\ directly for the first time with the physical values of $m_u$ and $m_d$, thereby removing this important uncertainty contribution~\cite{Chakraborty:2017tqp}.
To match the target experimental precision, however, the error on \amu\ must be further reduced to about 0.2\%.  

The $\order(\alpha^3)$ ``higher-order" hadronic vacuum-polarization contribution to \gmtwo\ is roughly 1.5\% that of the leading-order HVP contribution~\cite{Blum:2013xva}, and therefore only needs to be determined to around 10\% to match the projected experimental precision.  Experimental determinations from combining electron-positron inclusive scattering into hadrons data with dispersion relations quote errors of 0.4-0.9\%~\cite{Kurz:2015fhj,Jegerlehner:2017lbd,Keshavarzi:2018mgv}.  
Nevertheless, it is important to check these phenomenological values with {\it ab-inito} QCD calculations.  
Moreover, if the disagreement between theory and experiment persists or grows with the new Muon \gmtwo\ measurement, a complete first-principles Standard-Model theory value will be essential for drawing conclusions about the presence or nature of new physics.

In this paper we calculate the higher-order HVP contribution to \amu\ for the first time in lattice QCD.  
To enable us to focus on the methodology and error analysis, we use previously published lattice-QCD results for the Taylor coefficients of the renormalized vacuum polarization function ($\Pihat(Q^2)$) from Refs.~\cite{Donald:2012ga,Chakraborty:2014mwa,Colquhoun:2014ica,Chakraborty:2016mwy} to construct both \Pade~\cite{Chakraborty:2014mwa} and Mellin-Barnes approximants~\cite{Charles:2017snx} for $\Pihat(Q^2)$.  
Details on the lattice-QCD calculations can be found in these works.

This paper is organized as follows.  
In Sec.~\ref{sec:background}, we provide theoretical background on the hadronic-vacuum-polarization contributions to \gmtwo, and discuss our method for calculating the higher-order contributions.  
Next, in Sec.~\ref{sec:analysis} we present our analysis and error budget.  Last, in Sec.~\ref{sec:conclusions}, we show our final result  for $a_\mu^{\rm HVP,HO}$ and compare with non-lattice determinations.
Appendix~\ref{sec:appendix} provides expressions suitable for computing the $\order(\alpha^3)$ hadronic vacuum-polarization contribution to \amu\  directly from lattice-QCD simulations, while App.~\ref{sec:appendixMB} provides the definition of the $N=2+1+1$ Mellin-Barnes approximant for the $\Pihat(Q^2)$ used in this paper.  For completeness, App.~\ref{sec:TCs} gives the values of the quark-connected Taylor coefficients employed in our analysis.

\section{Theoretical background}
\label{sec:background}

\begin{figure}[tb]
\centering
\includegraphics[width=0.2\textwidth]{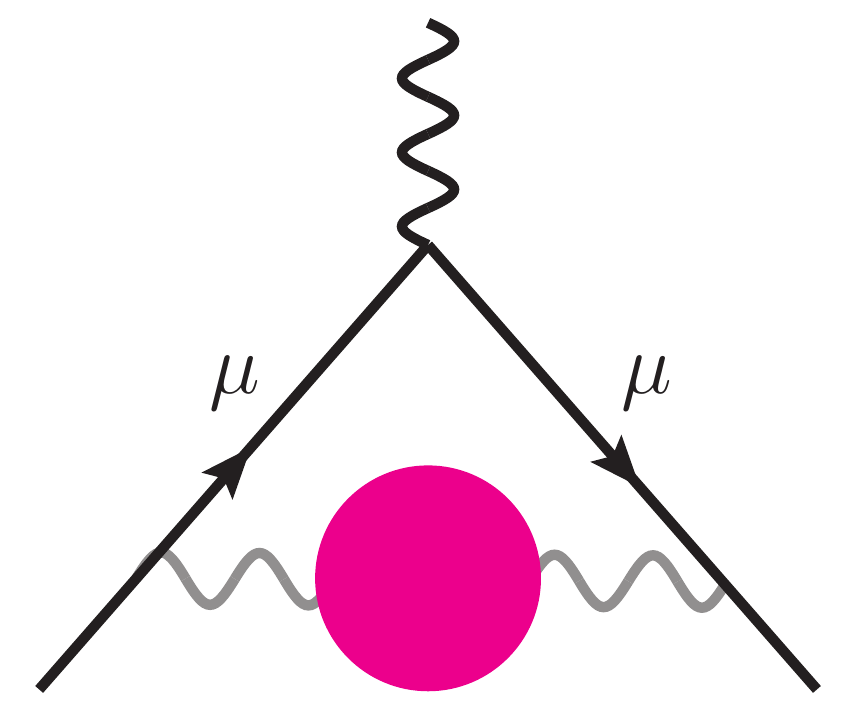}
\caption{
Leading hadronic contribution to the muon \gmtwo.  The shaded circle denotes all corrections to the internal photon propagator from the vacuum polarization of $u$, $d$, $s$, $c$, and $b$ quarks in the leading one-loop muon vertex diagram.}
\label{fig:HVP}
\end{figure}

\begin{figure*}[tb]
\centering
\includegraphics[width=0.19\textwidth]{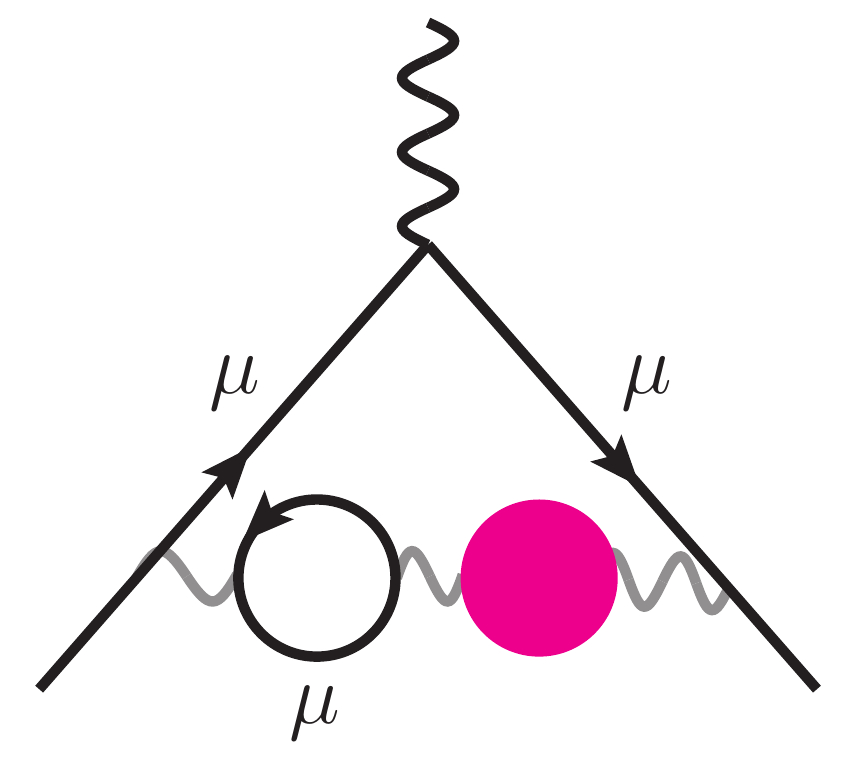} \includegraphics[width=0.19\textwidth]{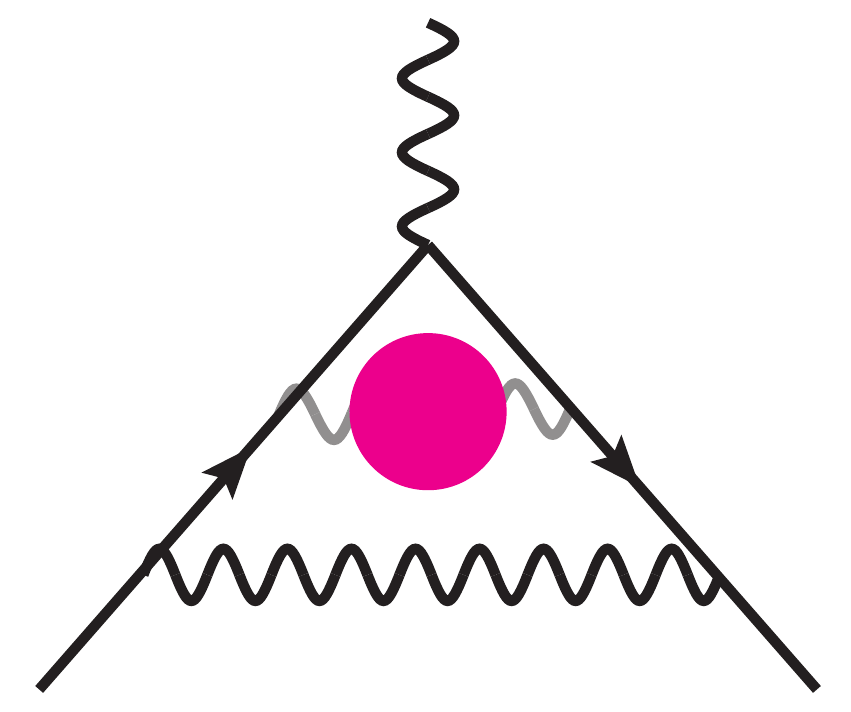} \includegraphics[width=0.19\textwidth]{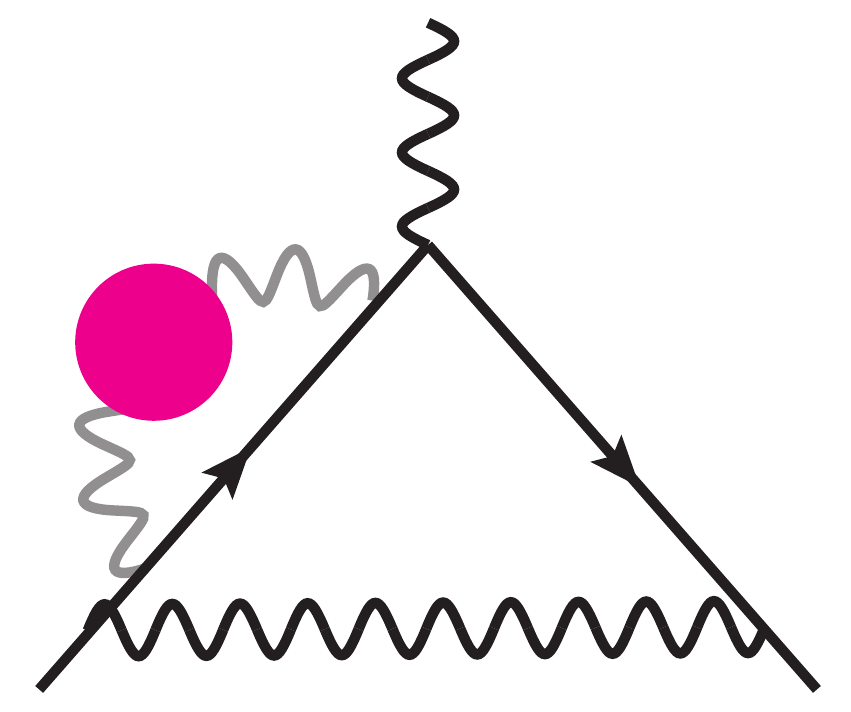} \includegraphics[width=0.19\textwidth]{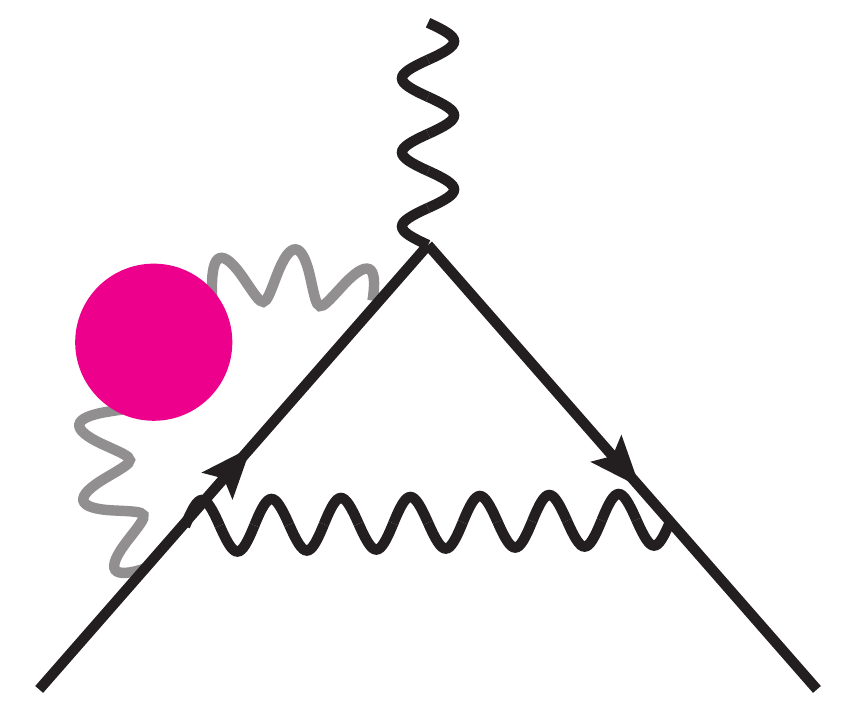} \includegraphics[width=0.19\textwidth]{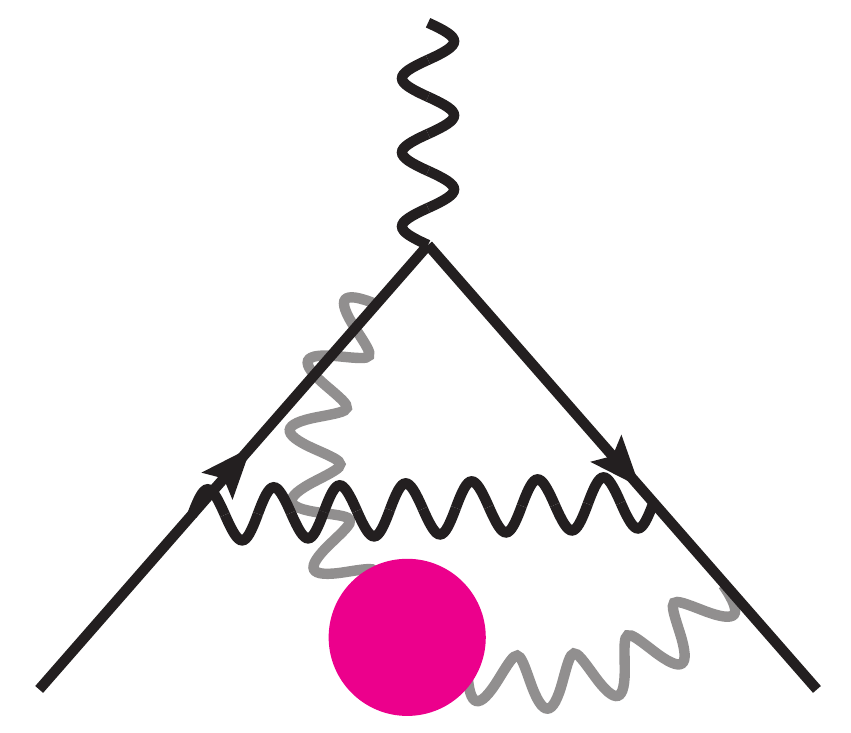} \\
(a) \\
\includegraphics[width=0.2\textwidth]{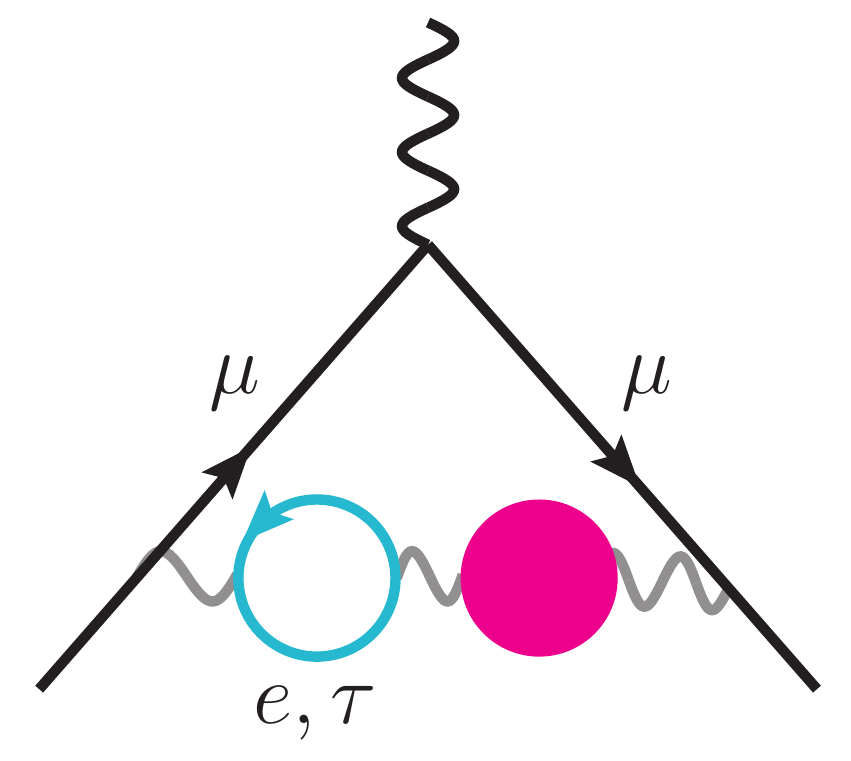} \qquad\qquad \includegraphics[width=0.2\textwidth]{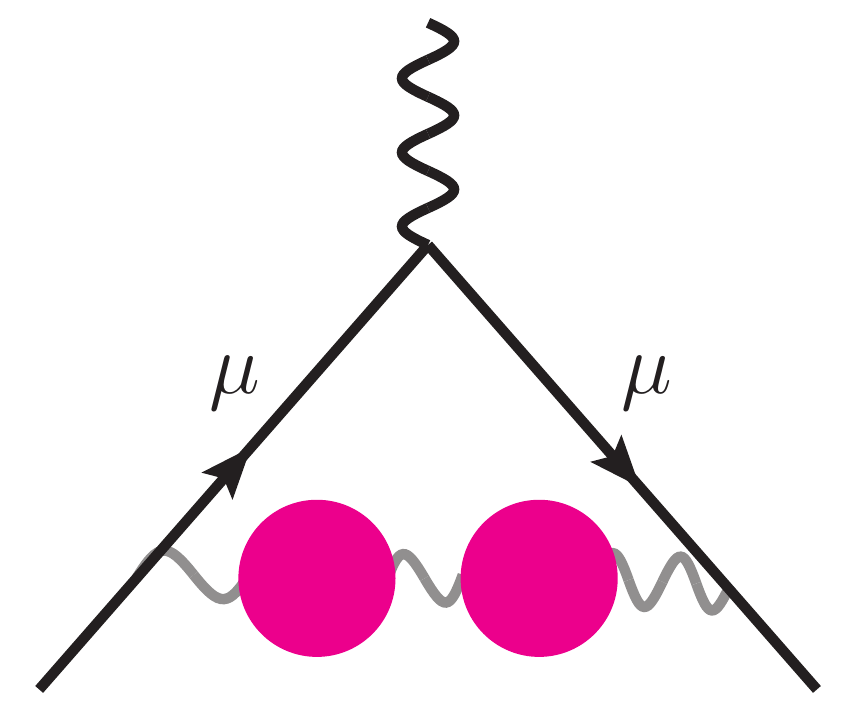} \\
(b) \qquad\qquad\qquad\qquad\quad (c)
\caption{Higher-order hadronic-vacuum-polarization contributions to \gmtwo.  For contribution (a),  diagrams that are reflections across the vertical axis through the center and diagrams in which the tree and corrected photon propagators are interchanged are not shown.
}
\label{fig:2abc}
\end{figure*}

The leading hadronic contribution to the muon anomalous magnetic moment arises from QCD corrections to the internal photon propagator in the $\order(\alpha^2)$ one-loop muon vertex diagram, as shown in Fig.~\ref{fig:HVP}.  At $\order(\alpha^3)$, higher-order hadronic contributions arise from adding a second internal photon line (as in Fig.~\ref{fig:2abc} (a)), adding a lepton loop to the existing photon line (as in Figs.~\ref{fig:2abc} (a) and (b)), or adding a second insertion of the hadronic vacuum polarization bubble on the photon line (as in Fig.~\ref{fig:2abc} (c)).  Both the leading- and NLO HVP contributions can be obtained, with the help of dispersion relations, from the energy scan of the experimental ``R-ratio"~\cite{Kurz:2015fhj,Jegerlehner:2017lbd,Davier:2017zfy,Keshavarzi:2018mgv}:
\begin{equation}
R_\gamma(s) \equiv \frac{\sigma(e^+ e^- \to \gamma^* \to {\rm hadrons})}{4\pi\alpha(s)^2 / (3s)} \,,
\label{eq:Rratio}
\end{equation}
where $s$ is the square of the center-of-mass energy. Table~\ref{tab:HVPRratio} shows two recent evaluations of the leading contribution and the individual higher-order contributions from diagrams (a), (b), and (c) by Jegerlehner~\cite{Jegerlehner:2017lbd} and Keshavarzi {\it et al.}~\cite{Keshavarzi:2018mgv}.  The higher-order contributions are roughly 1.5\% of the leading contribution, and do not contribute substantially to the total error on the Standard-Model theory value for $a_\mu$.

\begin{table}
\caption{
Determinations of the $\order(\alpha^2)$ (first column) and $\order(\alpha^3)$ hadronic-vacuum-polarization contributions (remaining columns) to \gmtwo\ from recent analysis of experimental data for the $e^+e^- \to {\rm hadrons}$ cross section by Jegerlehner~\cite{Jegerlehner:2017lbd} (top row) and Keshavarzi{\it et al.}~\cite{Keshavarzi:2018mgv} (bottom row).  \vspace{1mm}}
\label{tab:HVPRratio}
\begin{ruledtabular}
\begin{tabular}{lcccr}
\multicolumn{5}{c}{$10^{10}a_\mu^{\rm HVP}$}\\
Lowest order & (a) & (b) & (c) & total HO \\\hline
688.07(4.14) & -20.613(130) & 10.349(63) & 0.337(5) & -9.927(67) \\
693.27(2.46) & -20.77(8) & 10.62(4) &  0.34(1) & -9.82(4)
\end{tabular}
\end{ruledtabular}
\end{table}

Integrals for the $\order(\alpha^3)$ contributions from diagrams (a)--(c) have been presented in the literature in terms of $R_\gamma(s)$~\cite{Barbieri:1974nc,Krause:1996rf}.  These formulations, however, are not suited for our use, particularly in the case of contribution (a).  We therefore provide in Appendix~\ref{sec:appendix} new expressions for these contributions that are amenable to use with lattice-QCD data.  For each contribution, we provide two formulations to obtain $a_\mu^{(i)}; i=\{a,b,c\}$.   First, we use the following relationship between $R_\gamma(s)$ and the renormalized vacuum polarization function~\cite{Bernecker:2011gh},
\begin{equation}
\Pihat(q^2) = \frac{q^2}{3} \int_0^\infty ds\, \frac{R_\gamma(s)}{s(s+q^2)} \,, \label{eq:PihatToR}
\end{equation}
to derive expressions in terms of the renormalized vacuum polarization function $\Pihat(Q^2) \equiv \Pi(Q^2) - \Pi(0)$.\footnote{We use $q^2$ and $Q^2$ to denote the squared four-momenta in Minkowski and Euclidean space, respectively.}  These are the higher-order analogs of the original Blum formula for the leading HVP contribution~\cite{Blum:2002ii}, and are given in Eqs.~(\ref{eq:2aresult}), (\ref{eq:2bresult}), and~(\ref{eq:2cresult}).  We also provide expressions for the contributions from diagrams (a)--(c) directly in terms of the Euclidean vector-current correlator at zero momentum $G(t)$ using the relationship between $\Pihat(Q^2)$ and $G(t)$ below~\cite{Bernecker:2011gh}:
\begin{eqnarray}
\Pihat(\omega^2) &  \equiv & 4\pi^2 \left(\Pi(\omega^2) - \Pi(0)\right) \\
	& = & \frac{4\pi^2}{\omega^2} \int_0^{\infty} dt\, G(t) \left[ \omega^2 t^2 - 4 \sin^2\left(\frac{\omega t}{2}\right) \right]\,.\label{eq:PiHatLat}
\end{eqnarray}
These are the higher-order analogs of the time-momentum representation formulated by Bernecker and Meyer for the leading HVP contribution,  and are given in Eqs.~(\ref{eq:2aTMR}), (\ref{eq:2bTMR}), and~(\ref{eq:2cTMR}).

The higher-order HVP contributions are sensitive to the value of the renormalized vacuum polarization function at larger values of $Q^2$ than the leading-order contribution.   Figure~\ref{fig:HOintegrands}, left, plots the integrands for the leading-order and higher-order contributions as a function of $Q^2$ using the $N=2+1+1$ Mellin-Barnes approximant for $\Pihat(Q^2)$ from Ref.~\cite{Charles:2017snx}.  The integrand for the leading-order contribution is also shown for comparison.  The integrand of contribution (a) has large positive and negative contributions below $Q^2 = m_\mu^2$ that cancel substantially.  Because of this, the large-$Q^2$ region is numerically important, with about 5\% of the value of $a_\mu^{(a)}$ coming from $Q^2 > 10 {\rm GeV}^2$.  The integrand of contribution (b) peaks around $Q^2 = m_\mu^2 / 2\sqrt{2}$, and more than 95\% of the value of $a_\mu^{(b)}$ comes from $Q^2 < 0.5 {\rm GeV}^2$.  The integrand of contribution (c) peaks around $Q^2 = 2m_\mu^2$.  Because it is proportional to $\Pihat(Q^2)^2$, it decreases less rapidly with $Q^2$ than the other contributions; about 10\% of the value of $a_\mu^{(c)}$ comes from $Q^2 > 1 {\rm GeV}^2$.   Thus, it is important to employ approximants of $\Pihat(Q^2)$ that accurately reproduce the large-$Q^2$ behavior when calculating the higher-order contributions to \amu.

The higher-order HVP contributions are sensitive to the value of the Euclidean-time correlator at similar times as the leading-order contribution.   Figure~\ref{fig:HOintegrands}, right, plots the integrands for the leading-order and higher-order contributions~(a) and~(b) as a function of correlator time $t$ using $G(t)$ obtained from the spectral representation of $R_\gamma(s)$.  
(The kernel for contribution (c) depends upon the product of the correlator at two times $G(t) G(t^\prime)$ and thus the integrand cannot be conveyed in a one-dimensional plot.)
The leading-order (higher-order) kernels are proportional to $t$ ($t^2$) at small Euclidean times, and are proportional to $1/t$ (approach a constant )at large times, and the integrands all peak at around $t \sim 0.8$--1.0~fm.
The contributions to \amu\ from correlator data beyond 4~fm, which is approximately half the temporal extent (or less) of lattices employed in recent $g-2$ calculations, are about 0.5\% or less~\cite{Chakraborty:2016mwy,DellaMorte:2017dyu,Lehner:2017kuc,Borsanyi:2017zdw}.  

{\begin{figure*}[tb]
\centering
 \includegraphics[width=0.4\textwidth]{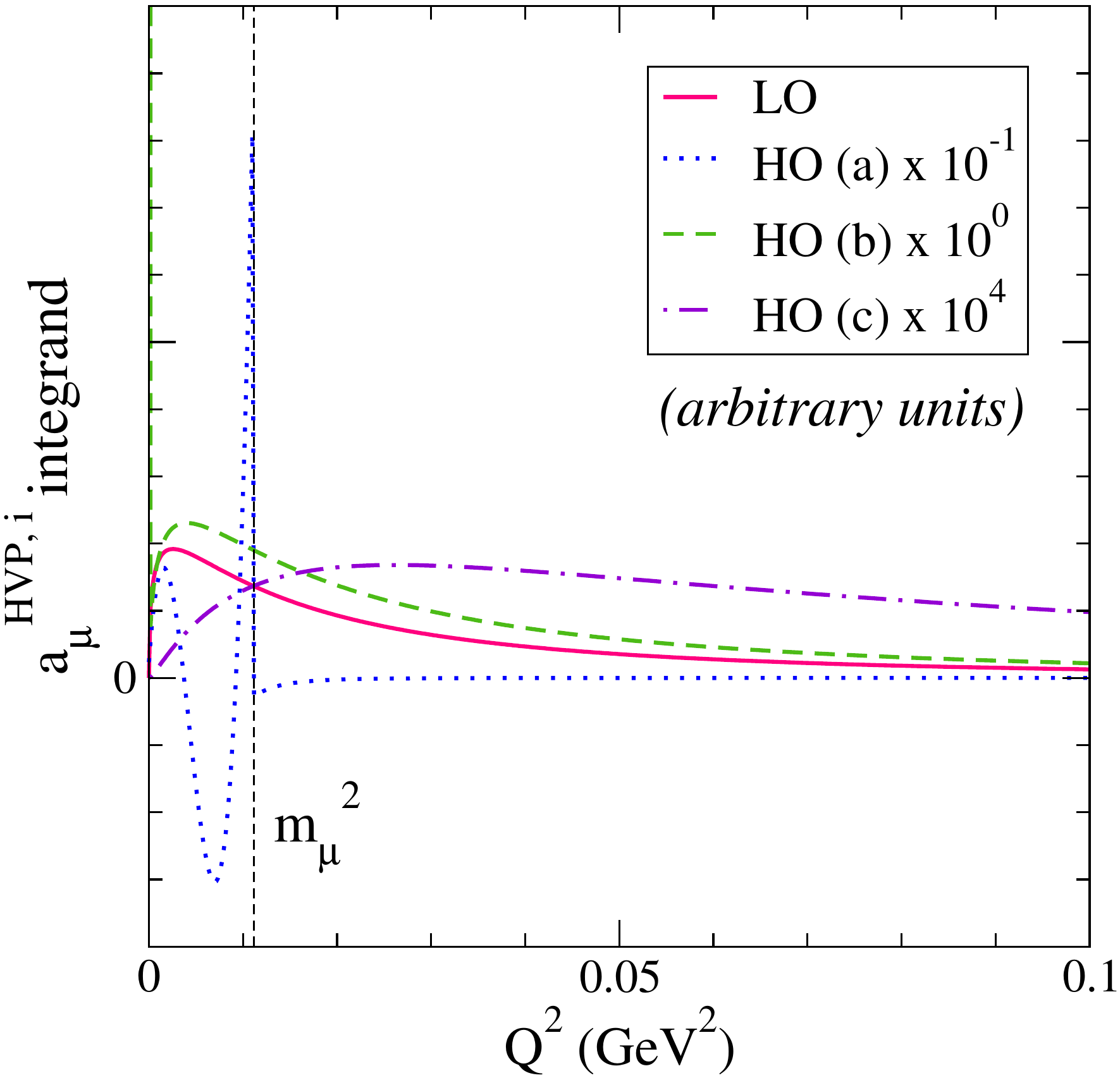} \qquad\qquad\qquad \includegraphics[width=0.4\textwidth]{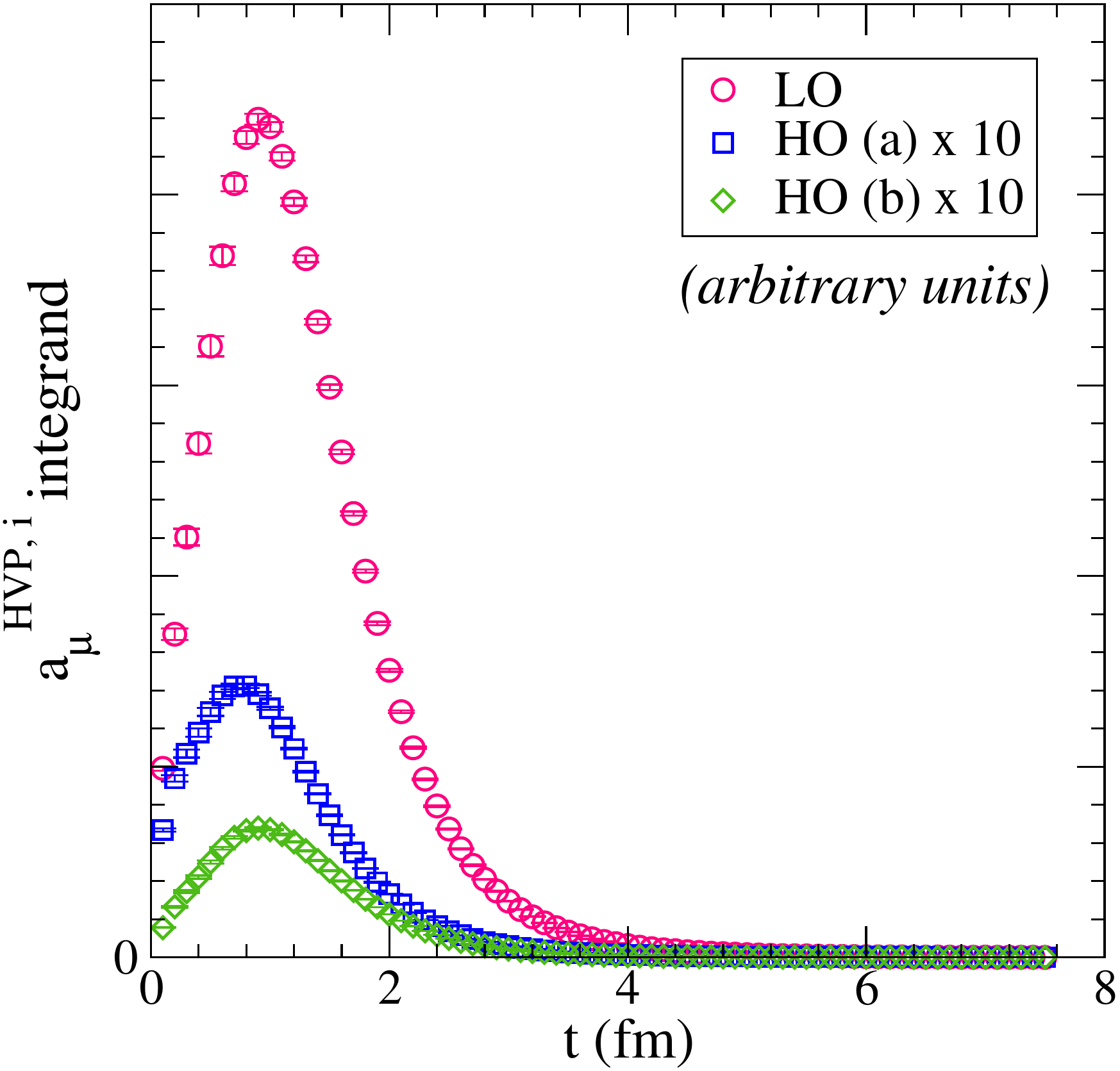} 
\caption{({\it color online.}) {\it Left:} integrands of Eqs.~(\ref{eq:2aresult}) (blue dots), (\ref{eq:2bresult}) (green dashes), and~(\ref{eq:2cresult}) (purple dot-dashes) obtained from the $N=2+1+1$ Mellin-Barnes approximant for $\Pihat(Q^2)$ given in Ref.~\cite{Charles:2017snx}, which employs preliminary moments of $R_\gamma(s)$ provided by Keshavarzi {\it et al.}~\cite{Keshavarzi:2018mgv}.  The leading-order integrand is also shown as a solid magenta line for comparison.  {\it Right:} integrands of Eqs.~(\ref{eq:2aTMR}) (blue squares) and~(\ref{eq:2bTMR}) (green diamonds) obtained from the parameterization of $R_\gamma(s)$ provided by Jegerlehner in his public \texttt{alphaQED FORTRAN} package~\cite{alphaQED}.  The leading-order integrand is also shown as magenta circles for comparison.}
\label{fig:HOintegrands}
\end{figure*}}

\section{Analysis}
\label{sec:analysis}

In this section we calculate the $\order(\alpha^3)$ contributions to \amu\ from the diagrams in Fig~\ref{fig:2abc}.
First, in Sec.~\ref{sec:method}, we describe the approximants of the renormalized vacuum function used to calculate the higher-order HVP contributions.
Next, we calculate the quark-connected contribution from light and heavy quarks in Sec.~\ref{sec:amu_ud}
Last, in Sec.~\ref{sec:amudisc}, we estimate the size of the quark-disconnected contribution.

\subsection{Approximants of $\Pihat(Q^2)$}
\label{sec:method}

We calculate the higher-order contributions to \amu\ using both \Pade\ and Mellin-Barnes approximants of the renormalized vacuum polarization function in the QED integrals given in Appendix~\ref{sec:appendix}.
Both approaches employ the Taylor coefficients $\Pi_i$ of $\Pihat(Q^2)$ expanded about $Q^2 = 0$:
\be
\Pihat(Q^2) = \sum_{i=1}^{\infty} \Pi_i Q^{2i}
\ee
As observed in Ref.~\cite{Chakraborty:2014mwa}, the $\Pi_i$ are proportional to the time-moments of the vector-current correlation function, and can be computed with small statistical errors in lattice QCD.  Further, with both the \Pade\ and Mellin-Barnes approches, only the first few Taylor coefficients are needed to obtain the leading-order HVP with a sub-percent systematic uncertainty associated with the parameterization of $\Pihat(Q^2)$~\cite{Chakraborty:2016mwy,Charles:2017snx}.

Following the method introduced by the HPQCD Collaboration~\cite{Chakraborty:2014mwa}, we construct the $[n,m]$ \Pade\ approximants for the renormalized hadronic vacuum polarization function from the $\Pi_i$'s.
The true result for $\Pihat(Q^2)$ is guaranteed to lie between the $[n,n]$ and $[n,n-1]$ \Pade\ approximants.
For the leading-order HVP contribution, the \Pade\ approximants provide a sufficiently accurate approximation of $\Pihat(Q^2)$ both at low and high $Q^2$ that the associated uncertainty in \amu\ is below 1\% by $n=2$~\cite{Chakraborty:2016mwy}.
Unfortunately, however, one cannot use the $[n,n-1]$ approximants $\Pihat(Q^2)$ to calculate the contributions to \amu\ from diagrams (a) and (c). This is because $\Pihat^{[n,n-1]}(Q^2) \sim Q^2$ as $Q^2 \to \infty$, making the integrals diverge in this limit.
The integrals using the $[n,n]$ \Pade\ approximants are well behaved, but another approach is needed to quantify the uncertainty in the higher-order contributions to \amu\ from the parameterization of $\Pihat(Q^2)$.

Recently de~Rafael and Charles {\it et al.} introduced the method of ``Mellin-Barnes approximants" to obtain \amu\ from the Taylor coefficients of $\Pihat(Q^2)$~\cite{deRafael:2017gay,Charles:2017snx}.
This approach uses the fact that the hadronic spectral function ${\rm Im}\Pihat(q^2)/\pi$ in QCD is positive and approaches a constant as $Q^2 \to \infty$ to identify a class of functions that can be employed as successive approximants to the Mellin transform ${\mathcal M}(s)$ of the hadronic spectral function.
Given $N$ moments of the Mellin transform ${\mathcal M}(-n)$, the Mellin-Barnes approximant ${\mathcal M}_{N}$ smoothly interpolates between these known values, and approaches the asymptotic value of ${\mathcal M}(s)$ from leading-order perturbative QCD as $s \to \infty$.  The Mellin moments are trivially related to the Taylor coefficients of $\Pihat(Q^2)$ as
\be
{\mathcal M}(-n)= 4\pi\alpha (-1)^n (4m_\pi^2)^{(n+1)} \Pi_{n+1} \,,
\ee
The first term in the moment expansion of the hadronic spectral function provides a rigorous upper bound on $\Pihat(Q^2)$ and \amu~\cite{Bell:1996md}.  In practice, the $N=1$ approximant obtained using ${\mathcal M}(0)$ from experimental $R_\gamma$ data yields a value for the leading-order HVP contribution that already agrees with the full result to better than 1\%~\cite{Charles:2017snx}.

\begin{figure}[tb]
\centering
\includegraphics[width=0.4\textwidth]{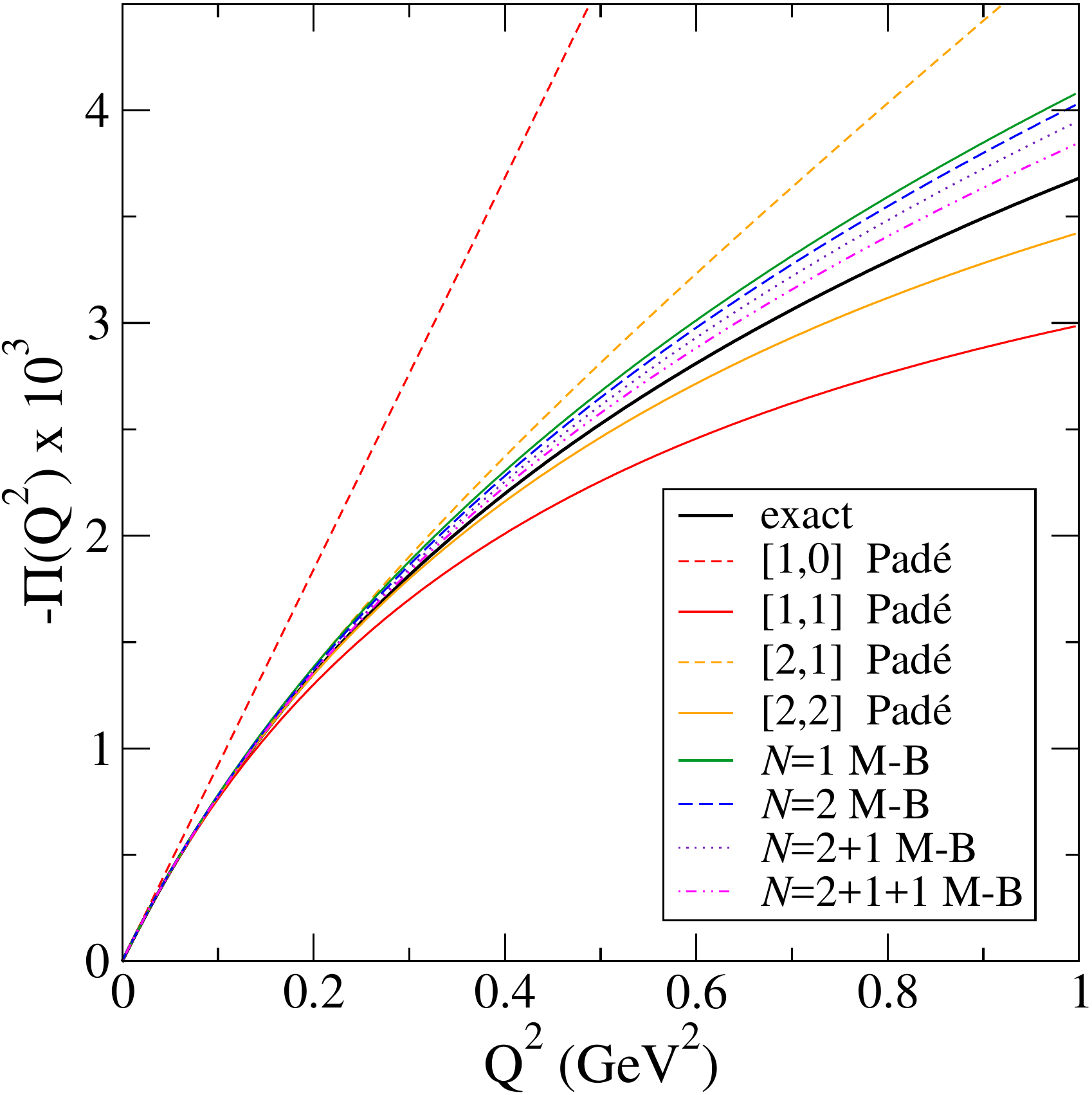}
\caption{({\it color online.}) First four \Pade\ approximants (``[1,0]--[2,2] \Pade") and Mellin-Barnes approximants (``N=1--N=2+1+1 M-B") of the renormalized vacuum polarization function calculated from the moments of $R_\gamma(s)$ analysis of Keshavarzi {\it et al.}~\cite{Keshavarzi:2018mgv}.  The exact result is shown as a solid black line for comparison~\cite{KeshavarzPrivateComm}.}
\label{fig:PihatApprox}
\end{figure}

Figure~\ref{fig:PihatApprox} plots the \Pade\ and Mellin-Barnes approximants for $\Pihat(Q^2)$ calculated from the first four moments of $R_\gamma(s)$~\cite{Keshavarzi:2018mgv}, and compares them with the exact result obtained from direct integration of $R_\gamma(s)$.
The Mellin-Barnes approximants are closer to the exact $\Pi(Q^2)$ than the \Pade s because they are constrained to satisfy the asymptotic perturbative-QCD behavior as $Q^2\to\infty$.  However, the rate at which the Mellin-Barnes approximants approach the true $\Pihat(Q^2)$ depends upon the specific functional form employed at each order.  In particular, the difference between successive approximants is not guaranteed to decrease with increasing $N$.  

As can be seen in Fig.~\ref{fig:PihatApprox}, for the realistic test case of the renormalized vacuum polarization function from experimental $R_\gamma(s)$ data, the \Pade\ and Mellin-Barnes approaches yield almost identical results at small $Q^2$.  For the numerically important region $Q^2 \leq 0.1$~GeV$^2$ shown in Fig.~\ref{fig:HOintegrands}, left, the [2,2] \Pade\ and 2+1+1 Mellin-Barnes approximants agree with each other -- and with the exact result -- to better than 0.15\%, which is within their statistical errors.  Further, when the approximants begin to diverge significantly at larger $Q^2$ values, the Mellin-Barnes approximants approach the exact $\Pihat(Q^2)$ from above, while the $[n,n]$ \Pade\ approximants approach it from below.  (The former is an empirical observation for the hadronic vacuum polarization in QCD~\cite{Charles:2017snx}, and not a generic property of Mellin-Barnes approximants.)  Consequently, the estimates of both the leading- and higher-order \amu\ obtained from the Mellin-Barnes and \Pade\ approximants bracket the exact results.  Therefore, for our numerical analysis in the following section, we take the average of the $\order(\alpha^3)$ contributions to \amu\ obtained from the 2+1+1 Mellin-Barnes and $[2,2]$ \Pade\ approximants for the central value, with error given by half the difference.  
This simple estimate is sufficiently accurate for illustrating our method for calculating the higher-order hadronic-vacuum-polarization contribution to the muon \gmtwo\ from lattice QCD.

\subsection{Quark-connected contribution}
\label{sec:amu_ud}

We calculate the $\order(\alpha^3)$ quark-connected contribution to \amu\ using the Taylor coefficients of $\Pihat(Q^2)$ obtained by the HPQCD Collaboration in Refs.~\cite{Donald:2012ga,Chakraborty:2014mwa,Colquhoun:2014ica,Chakraborty:2016mwy}.
The $u$, $d$, and $s$-quark Taylor coefficients were calculated on the MILC Collaboration's QCD four-flavor gauge-field configurations with highly-improved staggered (HISQ) sea and valence quarks~\cite{Follana:2006rc,Bazavov:2012xda}.  
The $b$-quark Taylor coefficients were also calculated on the HISQ ensembles, but with a radiatively-improved nonrelativistic QCD action for the $b$ quarks~\cite{Lepage:1992tx,Dowdall:2011wh}. 
The $c$-quark Taylor coefficients were calculated with HISQ valence quarks, but on MILC's three-flavor ensembles with asqtad sea quarks~\cite{Bernard:2001av,Aubin:2004wf,Bazavov:2009bb}.  
The MILC ensembles are isospin-symmetric, {\it i.e.} the up and down sea-quark masses are degenerate.  The light-quark mass varies from $m_l = m_s/5$ to Nature's value $m_l \sim m_s/27$, making a chiral extrapolation unnecessary, and the strange- (and charm-) sea-quark masses are fixed to close to their physical values.

We employ light- and strange-quark Taylor coefficients on two ensembles with physical light-quark masses and lattice spacings $a \approx 0.15$~fm and $0.12$~fm from Refs.~\cite{Chakraborty:2014mwa,Chakraborty:2016mwy}.  Table~\ref{tab:LQPis} gives the light- and strange-quark connected Taylor coefficients used in our analysis.  The values of $\Pi^{(ud)}_i$ include corrections for the finite lattice spatial volume and and nonzero lattice spacing computed at one-pion-loop order within scalar QED~\cite{Jegerlehner:2011ti}.   We employ charm- and bottom-quark Taylor coefficients from Refs.~\cite{Donald:2012ga,Colquhoun:2014ica}, which provide values of  $\Pi^{(c)}_i$ and $\Pi^{(b)}_i$ at the physical light-quark mass and in the continuum.  For convenience, Table~\ref{tab:Piscb} gives the heavy-quark connected Taylor coefficients used in our analysis.

To calculate the connected contribution to $a_\mu^{(\rm HO)}$, we first sum the individual Taylor coefficients $\Pi^{(ud)}_i$, $\Pi^{(s)}_i$, $\Pi^{(c)}_i$, and $\Pi^{(b)}_i$, and then use the total to construct the \Pade\ and Mellin-Barnes approximants for $\Pihat(Q^2)$.   
Beyond $N=2$, the functional forms of the Mellin-Barnes approximants are not unique; Appendix~\ref{sec:appendixMB} gives the form of $\Pihat_{2+1+1}(Q^2)$ used here.   We then use the resulting approximants for $\Pihat(Q^2)$ in the QED integrals, Eqs.~(\ref{eq:2aresult}), (\ref{eq:2bresult}), and~(\ref{eq:2cresult}), to obtain the quark-connected contributions to \amu\ from the diagrams in Fig.~\ref{fig:2abc}.     On each ensemble, and for each contribution (a)--(c), we average the values from the \Pade\ and Mellin-Barnes approximants, and take half the difference between the two as the systematic uncertainty from the parameterization of $\Pihat(Q^2)$.   Table~\ref{tab:amuHO} gives the results on the two ensembles employed in our analysis.

\begin{table}[tb]
    \caption{$\order(\alpha^3)$ hadronic-vacuum-polarization contributions to \gmtwo\ on two physical-mass HISQ ensembles obtained using [2,2] Pad{\'e} and $N=2+1+1$ Mellin-Barnes approximants for $\Pihat(Q^2)$. The uncertainties are from the errors on the Taylor coefficients and, for the averages, from the use of approximants for $\Pihat(Q^2)$.  \vspace{1mm}}
    \label{tab:amuHO}
\begin{ruledtabular}
\begin{tabular}{cccccc}
& \multicolumn{4}{c}{$10^{10}a_\mu^{{\rm HO,\, conn.}}$} \\
$\approx a$ (fm) & $\Pihat$ approx. & (a) & (b) & (c)  \\
\hline
0.15 & \Pade & -19.24(32) & 10.34(10) & 0.3186(79) \\
& M-B & -20.82(35) & 10.40(19) & 0.339(12) \\\hline
& Average & -20.03(82) & 10.37(11) & 0.329(12)  \\\hline
0.12 & \Pade & -19.05(29) & 10.176(87) & 0.3111(69) \\
& M-B & -20.58(27) & 10.23(15) & 0.3307(89) \\\hline
& Average & -19.82(79) & 10.204(91) & 0.321(11) 
\end{tabular}
\end{ruledtabular}
\end{table}

\begin{figure}[tb]
\centering
\includegraphics[width=0.4\textwidth]{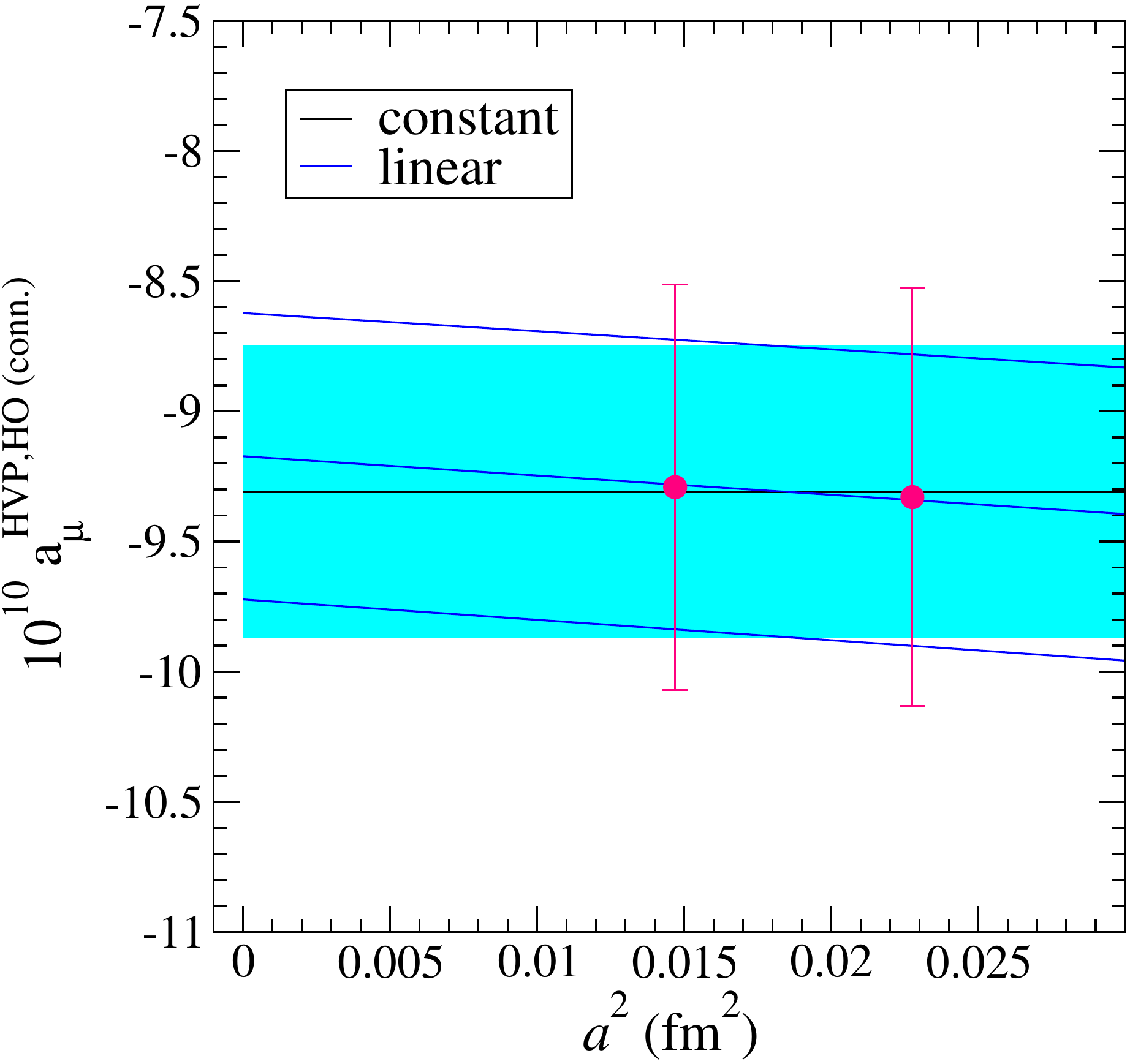}
\caption{({\it color online.}) Continuum extrapolation of $\order(\alpha^3)$ quark-connected contribution to \amu.  The filled cyan band shows the result of our preferred constant fit, while the solid blue lines show the result of a linear fit to Eq.~(\ref{eq:ContExtrap}) with the slope $c_{a^2}$ constrained with a Gaussian prior $0 \pm 1$.}
\label{fig:amuExtrap}
\end{figure}

Figure~\ref{fig:amuExtrap} shows the total $\order(\alpha^3)$ quark-connected contribution to \amu\ --- obtained by summing contributions (a)--(c)  in the rows labeled ``average" in Table~\ref{tab:amuHO} --- versus squared lattice spacing.    The data do not display any significant lattice-spacing dependence, so we fit them to constant to obtain the continuum-limit value of $a_\mu^{\rm HVP,HO}$.  We also consider an alternative linear extrapolation in $a^2$ to a function of the form
\be
a_\mu^{\rm HVP,HO} \left( 1 + c_{a^2} \frac{(a\Lambda)^2}{\pi^2} \right), \label{eq:ContExtrap}
\ee
with $\Lambda = 500$~GeV a typical QCD scale.  The linear-fit result for $c_{a^2}$ is consistent with zero, and for $a_\mu^{\rm HVP,HO}$ is close to the value from the constant fit.  We therefore conclude that discretization effects are smaller than the fit error on $a_\mu^{\rm HVP,HO}$, and do not assign a separate systematic error from this source.

The HPQCD Collaboration reduced the statistical errors in the light-quark connected Taylor coefficients in Ref.~\cite{Chakraborty:2016mwy} by using fit results for the vector-current correlators for times greater than 1.5~fm.  Although the lowest-energy states in these correlators are $I=1\  \pi\pi$ pairs, no evidence of such states was seen in the two-point fits, and the ground-state energies obtained are consistent with the experimental $\rho^0$ meson mass.
HPQCD estimate the contribution to the leading-order light-quark connected contribution to \amu\ from the omitted $\pi\pi$ states within scalar QED to be $3 \times 10^{-10}$.  
We expect $\pi\pi$ contributions to be similar in size for the dominant higher-order diagrams (a) and (b) because the integrands in Eqs.~(\ref{eq:2aresult}) and~(\ref{eq:2bresult}) are proportional to $\Pihat(Q^2)$, just as for the leading-order hadronic vacuum polarization.
Hence, we take the same percentage error of 0.5\% as the uncertainty in $a_\mu^{\rm HVP,HO}$ from $\pi\pi$ states below the $\rho$ pole.  

The four-flavor gauge-field ensembles employed in our analysis have degenerate up and down sea-quark masses.
Recently the Fermilab Lattice, HPQCD, and MILC Collaborations calculated the strong-isospin-breaking correction to \amu\ for the first time with physical values $m_u$ and $m_d$~\cite{Chakraborty:2017tqp}.  
They obtain +1.5(7)\% for the relative correction that should be applied to the leading-order light-quark connected contribution, in agreement with phenomenological estimates~\cite{Wolfe:2010gf,Jegerlehner:2011ti,Jegerlehner:2017gek}.
Here we use +1.5(1.0)\% to correct the continuum-limit value of $a_\mu^{\rm HVP,HO}$ from Fig.~\ref{fig:amuExtrap}, where we have taken a larger uncertainty of 1\% on the relative correction to account for the fact that the shift was not calculated directly for the higher-order hadronic vacuum polarization. 

The QCD gauge-field ensembles employed in our analysis do not include effects due to the quarks' nonzero electromagnetic charges in Nature.  
The dominant QED effect in \amu\ arises from producing a hadron polarization bubble consisting of a $\pi^0$-$\gamma$ pair.
Following Hagiwara {\it et al.}~\cite{Hagiwara:2003da} we calculate the contribution to $a_\mu^{\rm HVP,HO}$ from $e^+e^- \to \pi^0\gamma$ in the region $0.6 < \sqrt{s} < 1.03$~GeV using the latest experimental data for this channel from the SND Experiment~\cite{Achasov:2016bfr}.  We obtain 
\be
\Delta a_\mu^{({\rm HO},\pi^0\gamma)} = -0.056(8) \times 10^{-10} \,,
\ee
which is approximately 0.6\% of the total quark connected contribution.  We therefore take 1\% as the error from the omission of electromagnetism in the simulations.  

Finally, as discussed in Appendix~\ref{sec:appendix}, in order to express higher-order contribution 2(a) in Fig.~\ref{fig:HVP} in terms of the renormalized vacuum polarization function, we must drop terms in the original integrand~\cite{Barbieri:1974nc,Krause:1996rf} that are proportional to $(m_\mu^2/s)^n\log^2(m_\mu^2/s)$.   We have calculated the numerical size of these terms from experimental $R_\gamma(s)$ data~\cite{alphaQED} and, although they are small, they are not negligible given the size of our statistical and other systematic uncertainties.  
To account for the omission of the ``log$^2$'' in our calculation of contribution 2(a) via Eq.~(\ref{eq:2aresult}), we therefore include an additional systematic uncertainty of $1 \times 10^{-10}$, which is almost twice the size of these terms calculated from $R_\gamma(s)$ data.

Table~\ref{tab:amuErrBudget} gives the complete error budget for the $\order(\alpha^3)$ quark-connected contribution to \amu.  
The largest uncertainties are associated with the omitted ``log$^2$'' terms in contribution 2(a) and from the use of \Pade\ and Mellin-Barnes approximants for the renormalized vacuum polarization function.
Although the estimated uncertainties from the omission of QED and isospin breaking in the gauge-field configurations, and from low-lying $\pi\pi$ states in the vector-current correlators, are based on calculations for the leading-order vacuum polarization, they are about four times smaller, and do not contribute substantially to the total error.
We obtain for the quark-connected contribution to $a_\mu^{\rm HVP,HO}$ with all systematics included
\be
\!\!\!\!\!\!\!\!10^{10} a_\mu^{({\rm HO, conn.})} =  -9.45(18)_{\rm lat.} (55)_{\widehat\Pi-{\rm approx.}}  (1.0)_{\log^2} \,, \label{eq:amu_conn}
\ee
where ``lat." denotes the sum of contributions associated with the underlying lattice-QCD calculations of the Taylor coefficients.

\begin{table}[tb]
    \caption{Error budget for $\order(\alpha^3)$ quark-connected contribution to \gmtwo.  \vspace{1mm}}
    \label{tab:amuErrBudget}
\begin{ruledtabular}
\begin{tabular}{lr} 
 & $a_\mu^{({\rm HO},ud)}$ (\%) \\ 
\hline
Omission of log$^2$ terms & 10.6 \\ 
Pad\'e approximants & 5.8 \\ 
Isospin-breaking and electromagnetism & 1.4 \\ 
Taylor coefficients & 1.2 \\ 
$\pi\pi$ states ($t^*$) & 0.5 \\ 
\hline 
Total & 12.2 \\ 
\end{tabular}
\end{ruledtabular}
\end{table}

\subsection{Quark-disconnected contribution}
\label{sec:amudisc}

Although several lattice-QCD calculations of the leading-order quark-disconnected contribution to \amu\ are available~\cite{Chakraborty:2015ugp,Blum:2015you,Borsanyi:2017zdw}, these publications do not provide the Taylor coefficients of the renormalized vacuum polarization function.\footnote{In Ref.~\cite{Borsanyi:2016lpl}, the BMW Collaboration provides the first two Taylor coefficients $\Pi_1^{({\rm disc.})}$ and $\Pi_2^{({\rm disc.})}$, which are not sufficient to construct the [2,2] \Pade\ and $N=2+1+1$ Mellin-Barnes approximants.}
We therefore estimate the values of the quark-disconnected Taylor coefficients assuming ground-state dominance of the vector-current correlators as in Ref.~\cite{Chakraborty:2015ugp}.  Using Eq.~(11) of that work,
\be
\frac{Q^2\Pi_i^{(\rm disc.)}}{Q^2\Pi_i^{({\rm conn.})}} = \frac{1}{10} \left[\frac{m_\rho^{2j+2}f_\omega^2}{m_\omega^{2j+2}f_\rho^2} - 1 \right] \,, \label{eq:PiDisc}
\ee
with $\{M_\rho, M_\omega\} = \{ 0.77526(25), 0.78265(12)\}$~GeV from the PDG~\cite{Olive:2016xmw} and $\{f_\rho, f_\omega\} = \{0.21(1),0.20(1)\}$~GeV yields
\be
{Q^2\Pi_1^{(\rm disc.)}}/{Q^2\Pi_1^{({\rm conn.})}} = -0.013(12) \,, \label{eq:amu_disc}
\ee
and similar results for the higher Taylor coefficients.
Both the leading $\order(\alpha^2)$ contribution to \amu\ and the domiant $\order(\alpha^3)$ contributions from diagrams (a) and (b) are proportional to the Taylor coefficient $\Pi_1$ at lowest order in the small-$Q^2$ expansion.  Further, the dominant quark-connected contribution is from the light up and down quarks.
We therefore take $-1.3(1.2)\%$ as the correction and uncertainty due to the omission of quark-disconnected contributions in our analysis.
We note that our estimate in Eq~(\ref{eq:amu_disc}) is consistent with recent lattice-QCD calculations of the leading-order quark-disconnected contribution with physical-mass pions from the BMW~\cite{Borsanyi:2017zdw} and RBC/UKQCD Collaborations~\cite{Blum:2015you}, who obtain for the ratio ${a_\mu^{({\rm LO, disc.})}}/{a_\mu^{({\rm LO}, u/d\ {\rm conn.})}}$ approximately -2.0\% and -1.5\%, respectively.

\section{Result and outlook}
\label{sec:conclusions}

The $\order(\alphaEM^3)$ hadronic-vacuum polarization contribution is a necessary ingredient in an {\it ab-initio}-QCD determination of the hadronic contributions to \gmtwo.  Towards this aim, we have introduced a new method for calculating the higher-order HVP contribution from lattice QCD, deriving formulae in terms of either the Euclidean vector-current correlator or the renormalized vacuum polarization function.  These are given in Appendix~\ref{sec:appendix}, and are the key results of this work.

We demonstrate the approach using the Taylor coefficents of the renormalized vacuum polarization function at the physical light-quark mass and two lattice spacings from Ref.~\cite{Chakraborty:2016mwy}. The total higher-order hadronic vacuum polarization contribution to \gmtwo\ is obtained by adding our calculation of the quark-connected contribution, Eq.~(\ref{eq:amu_conn}), to our estimate of the quark-disconnected contribution, Eq.~(\ref{eq:amu_disc}).  Our final result is
\be
10^{10} a_\mu^{\rm HVP,HO} = -9.3(0.6)_{\rm conn.} (0.1)_{\rm disc.} (1.0)_{{\rm log}^2} \,, \label{eq:amu_result}
\ee
where the first two errors errors are from the quark-connected and quark-disconnected contributions, respectively.  We list the error from  omission of the ``log$^2$" terms separately, since it does not arise from the use of lattice QCD to obtain the renormalized vacuum polarization function.   This error could be eliminated with a different trick for expressing contribution (a) in terms of $\hat\Pi(Q^2)$ than the one employed here.   Equation~(\ref{eq:amu_result}) is the first lattice-QCD determination of the higher-order hadronic vacuum polarization contribution to \gmtwo.  It is consistent with determinations from  $e^+ e^- \to {\rm hadrons}$ data~\cite{Kurz:2015fhj,Jegerlehner:2017lbd,Keshavarzi:2018mgv}, but with an approximately ten times larger error.  

A significant -- and difficult to quantify -- uncertainty in Eq.~(\ref{eq:amu_result}) stems from our use of approximants for the renormalized vacuum polarization function, which we employ so that we can exploit already-published values of the Taylor coefficients.  Our estimated error covers the results for $a_\mu^{\rm HVP,HO}$ from both the \Pade\ and Mellin-Barnes approaches; this is based on the empirical observation that the exact result for $\hat\Pi(Q^2)$ obtained from $R_\gamma(s)$ lies between the two types of approximants.
Fortunately, this error can be eliminated by calculating the $\order(\alphaEM^3)$ contributions directly from the lattice vector-current correlators.   
We will update our initial results using this theoretically cleaner approach, and also analyze ensembles with finer lattice spacings, in a future work.

Confirmation from independent lattice-QCD calculations is also essential before any results can be combined with experimental $g-2$ measurements to test the Standard Model.  The tools developed in this paper will enable others to provide this.

\acknowledgements
We thank John Campbell, J{\'e}r{\^o}me Charles, David Greynat, Fred Jegerlehner, Eduardo de~Rafael, and Thomas Teubner for valuable discussions; Alex Keshavarzi for providing additional information on the KNT R-ratio analysis; and David Greynat for comments on the manuscript.
We also thank our colleagues in the Fermilab Lattice and MILC Collaborations for providing the gauge-field configurations employed in this work, for invaluable contributions to other components of our multi-year project to calculate the hadronic vacuum polarization contribution to the muon $g-2$, and for general support. 
The calculation in this work was inspired by discussions at the first plenary workshop of the Muon $g-2$ Theory Initiative, and benefitted from interactions at subsequent meetings.

The computations discussed here were  carried out on the Darwin Supercomputer at the DiRAC facility, which is jointly funded by the U.K.\ Science and Technology Facility Council, the U.K.\ Department for Business, Innovation and Skills, and the Universities of Cambridge and Glasgow.   

This work was supported in part by the U.S.\ Department of Energy under grant PHY13-16222 (G.P.L.) and by the Gilmour bequest to the University of  Glasgow and the STFC (C.T.H.D.).
Fermilab is operated by Fermi Research Alliance, LLC, under Contract No.\ DE-AC02-07CH11359 with the United States Department of
Energy, Office of Science, Office of High Energy Physics.
The United States Government retains and the publisher, by accepting the article for publication, acknowledges that the United
States Government retains a non-exclusive, paid-up, irrevocable, world-wide license to publish or reproduce the published form of this manuscript, or allow others to do so, for United States Government purposes.

\appendix
\section{Formulae for higher-order HVP contributions to \gmtwo}
\label{sec:appendix}

Here we present integrals that can be used to calculate the higher-order hadronic-vacuum-polarization contributions
to \gmtwo\ from lattice-QCD data.  Our starting point is the expressions derived by Krause in Ref.~\cite{Krause:1996rf} for the contributions from diagrams (a)--(c) in Fig.~\ref{fig:2abc} in terms of  $R_\gamma(s)$ [Eq.~(\ref{eq:Rratio})].   Contributions (b) and (c) can be expressed as the 1-loop QED integral for the lowest-order contribution from Blum~\cite{Blum:2002ii} with a simple replacement of $\hat \Pi(Q^2)$, whereas contribution (a) is a nontrivial result of this work. 

\subsection{Contribution (a)}

A complete analytical result for the contribution from the diagrams in (a) of Fig.~\ref{fig:2abc} was first presented by Barbieri and Remiddi in Ref.~\cite{Barbieri:1974nc}; in this work they also provide an expansion to first order in $m_\mu^2/s$.   Later, in Ref.~\cite{Krause:1996rf}, Krause derived an asymptotic expansion for the kernel function in terms of the parameter $r = m_\mu^2/s$, which is more amenable to numerical integration.  We start with the asymptotic expression given in Eq.~(7) of Krause, which contains powers and logarithms of $r$.

Equation~(7) does not have the form needed to exploit the relationship between $R_\gamma(s)$ and the renormalized vacuum polarization function in Eq.~(\ref{eq:PihatToR}).
As suggested by Groote {\it et al.}~\cite{Groote:2001vu}, however, one can exploit generating integral representations of $r^n$ and $r^n$log($r$) to express the pure polynomial and log terms in the asymptotic expansion of the kernel function in terms of $\Pihat$.  Using Eqs.~(39)--(42) of that work, and discarding terms proportional to $\log^2(r)$ yields the following integral expression for contribution $(a)$ in terms of the renormalized vacuum polarization function:
\begin{eqnarray}
a_\mu^{(a)}&=& \left( \frac{\alpha}{\pi} \right)^3 \int_0^1 dx \, \left[
 (a_0 + a_1 x + a_2 x^2 + a_3 x^3) \, \Pihat\left( \frac{m_\mu^2}{x} \right) \right. \nonumber \\
& + & \left. \frac{(b_0 + b_1 x + b_2 x^2 + b_3 x^3)}{x} \, \Pihat\left( m_\mu^2 x \right)  \right] \,,
\label{eq:2aresult}
\end{eqnarray}
with
\begin{eqnarray} &&
a_0 = -\frac{23}{18}\,, \quad
b_0 = \frac{61791297-7818200 \pi ^2}{1200} \,, \quad \nonumber \\ &&
a_1 = \frac{367}{108} \,, \quad
b_1 = -\frac{724746871}{1200}+\frac{152879 \pi ^2}{2} \,, \nonumber \quad \\ &&
a_2 = -\frac{10079}{1800} \,, \quad
b_2 = \frac{5364282053}{3600}-\frac{377219 \pi ^2}{2} \,, \quad \\ &&
a_3 = \frac{6517}{900} \,, \quad
b_3 = -\frac{70906297}{72}+\frac{373975 \pi ^2}{3} \,. \nonumber
\end{eqnarray}
Checking the size of the omitted logarithmic terms using experimental data for $R_\gamma(s)$~\cite{alphaQED}, we find that they are below $1 \times 10^{-10}$.  

Alternatively, contribution (a) is given in terms of the Euclidean zero-momentum correlator by
\begin{eqnarray}
a_\mu^{(a)} & = & \frac{4 \alpha^3}{\pi} \int_0^{\infty} dt\,  t^2\, G(t) \;\tilde{K}^{(a)}_\ell(t) \,,
\label{eq:2aTMR}
\end{eqnarray}
with
\begin{widetext}
\begin{eqnarray}
\tilde{K}^{(a)}_\ell(t) &=& \frac{1}{t^2} \int_0^1 dx   \left\{ \frac{1}{\omega^2}  \sum_{i=0}^3 a_i x^{i}   \left[ \omega^2 t^2 - 4\sin^2\left(\frac{\omega t}{2}\right) \right] +  \frac{1}{{\omega^\prime}^2 x}  \sum_{i=0}^3 b_i x^{i} 
 \left[ {\omega^\prime}^2 t^2 - 4\sin^2\left(\frac{\omega^\prime t}{2}\right) \right] \right\} \,,
\label{eq:KaTilde}
\end{eqnarray}
\end{widetext}
and
\begin{equation}
\omega^2 = \frac{m_\mu^2}{x} \,, \quad {\omega^\prime}^2 = m_\mu^2 x \,. 
\end{equation}
The factors of $t^2$ and $1/t^2$ in Eqs.~(\ref{eq:2aTMR}) and~(\ref{eq:KaTilde}), respectively, are chosen to make the kernel function $\tilde{K}^{(a)}(t)$ dimensionless.  With these formulae, contribution (a) can be obtained from a simple weighted sum of $G(t)$ as in the leading-order case.

\subsection{Contribution (b)}

We start from Eq.~(9) of Ref.~\cite{Krause:1996rf} and make the change of variables $Q^2 = m_\mu^2 x^2/(1-x)$.
The contribution from diagram (b) in Fig.~\ref{fig:2abc} is then given in terms of the renormalized vacuum polarization function by 

\begin{equation}
a_\mu^{(b)} = 8\pi^2 \left( \frac{\alpha}{\pi} \right)^3 \int_0^\infty dQ^2 K_E(Q^2) 
\Pihat\left( Q^2 \right) F^{\ell} \left(m_e^2, Q^2\right),
\label{eq:2bresult}
\end{equation}
where the lepton loop function is 
\begin{align}
F^{\ell} \left(m_e^2, x\right) 
& = -\frac{8}{9} + \frac{\beta^3}{3} - \left( \frac{1}{2} - \frac{\beta^2}{6} \right) \beta \log \left( \frac{\beta - 1}{\beta+1} \right) \,,\\
\beta & \equiv  \sqrt{ 1 + 4 \left( \frac{m_e^2}{Q^2} \right)} \,.
\end{align}
and 
$K_E(Q^2)$ is the standard kernel function introduced by Blum in Ref.~\cite{Blum:2002ii}:
\bea
  \label{eq:kerK}
K_E(Q^2) &=& \frac{1}{m_\mu^2}\cdot \hat s\cdot Z(\hat s)^3\cdot 
\frac{1 - \hat s Z(\hat s)}{1 + \hat s Z(\hat s)^2}\,,
\\
Z(\hat s) &=& - \frac{\hat s - \sqrt{\hat s^2 + 4 \hat s}}{2  \hat s},
\quad \hat s = \frac{Q^2}{m_\mu^2}\,.
\eea
Thus, the expression in Eq.~(\ref{eq:2bresult}) is simply the leading-order QED integral with the replacement $\hat\Pi(Q^2) \to 8\pi\alpha \times \Pihat\left( Q^2 \right) F^{\ell} \left(m_e^2, Q^2\right)$.
The analogous contribution from the $\tau$ lepton is negligible because it is suppressed by $m_\mu^2/m_\tau^2$.

Contribution (b) can also be obtained from a weighted sum of the Euclidean zero-momentum correlator as in the leading-order case:
\begin{eqnarray}
a_\mu^{(b)}(m_\ell) = \frac{8 \alpha^3}{\pi}  \int_0^{\infty} dt\, t^2\, G(t) \;\tilde{K}^{(b)}_\ell(t; m_\ell) \,,
\label{eq:2bTMR}
\end{eqnarray}
with the dimensionless kernel
\begin{widetext}
\begin{equation}
\tilde{K}^{(b)}_\ell(t; m_\ell) = \frac{1}{t^2} \int_0^\infty  d\omega \frac{4\pi^2 K_E(\omega^2)}{\omega^2}  \left[ \omega^2 t^2 - 4\sin^2\left(\frac{\omega t}{2}\right) \right] \,F^\ell \left(m_\ell^2, \omega^2\right) \,.
\end{equation}
\end{widetext}

\subsection{Contribution (c)}

We start from Eq.~(13) of Ref.~\cite{Krause:1996rf}.  Diagram (c) in Fig.~\ref{fig:2abc} contains two hadronic insertions, and thus the contribution depends upon the square of the renormalized vacuum polarization function:
\begin{eqnarray}
a_\mu^{(c)} &=& 4\pi^2 \left( \frac{\alpha}{\pi} \right)^3 \int_0^\infty dQ^2 K_E(Q^2)  \Pihat(Q^2)^2 \,.
\label{eq:2cresult}
\end{eqnarray}
In this case, the expression in Eq.~(\ref{eq:2cresult}) has the form of the 1-loop QED integral, but with the replacement $\hat\Pi(Q^2) \to 4\pi\alpha \times \Pihat(Q^2)^2$. 

When contribution (c) is expressed in terms of the Euclidean zero-momentum correlator, the two powers of the vacuum polarization function above yield two integrals over times $t$ and $t^\prime$:
\begin{equation}
a_\mu^{(c)}   =  16 \pi \alpha^3  \int_0^{\infty} \!\!\!dt\, t^2\, G(t) \int_0^{\infty} \!\!\!dt^\prime\, {t^\prime}^2 G(t^\prime) \;\tilde{K}^{(c)}(t,t^\prime) \,,
\label{eq:2cTMR}
\end{equation}
with the dimensionless kernel 
\begin{widetext}
\begin{equation}
\tilde{K}^{(c)}(t,t^\prime) = \frac{1}{t^2 {t^\prime}^2} \int_0^\infty  d\omega \frac{4\pi^2 K_E(\omega^2)}{\omega^2} \left[ \omega^2 t^2 - \sin^2\left(\frac{\omega t}{2}\right) \right]  \bigg[ \omega^2 t^{\prime 2} - \sin^2\bigg(\frac{\omega t^\prime}{2}\bigg) \bigg] \,.
\end{equation}
\end{widetext}
This formulation is slower to implement numerically than the analogous formulae for contributions (a) and (b) due to the double integral.

\section{Definition of $\Pihat_{2+1+1}(Q^2)$}
\label{sec:appendixMB}

In this paper we employ a slightly different form for the $N=2+1+1$ approximant for the Mellin transform of the hadronic spectral function than of the one given in Ref.~\cite{Charles:2017snx}, using
\begin{align}
\!\!\!\!{\mathcal M}_{2+1+1}(s) &= \frac{\alpha \sum_f Q_f^2}{3\pi} \left\{\frac{1}{1-s}\frac{\Gamma(a-s)\Gamma(b-1)}{\Gamma(a-1)\Gamma(b-s)} + \right. \nonumber\\ & \left. \Gamma(1-s)\frac{\Gamma(c - 1)}{\Gamma(c - s)} + \Gamma(1-s)\frac{\Gamma(d - 1)}{\Gamma(d - s)} \right\}\,,
\end{align}
with $Q_f$ the charge of each quark flavor in units of $e$.  We obtain the coefficients $a$--$d$ by solving the matching conditions 
\begin{align}
{\mathcal M}_{2+1+1}(-n) = {\mathcal M}_{\rm LQCD}(-n)\,, \quad n = \{0,1,2,3\} \,,
\end{align}
where ${\mathcal M}_{\rm LQCD}(-n)$ are the lattice Mellin moments, and choosing the solution that satisfies ${\rm Re}(a,b,c,d) \geq 1$, ${\rm Im}(a,b)=0$, and $c = d^*$.   The corresponding approximant for $\Pihat(Q^2)$ is then given by the following sum of generalized hypergeometric functions:
\begin{align}
\Pihat_{2+1+1}(Q^2) =  \frac{\alpha \sum_f Q_f^2}{\pi} z \left\{ 
\frac{(a-1)}{(b-1)} \ _{3}{F}_{2}\left[ \begin{array}{ccc} 1 & 1 & a \\ ~ & 2 & b\end{array}; {-z}\right] \right. \nonumber\\ 
+ \left. \frac{1}{(c-1)} \ _{2}{F}_{1}\left[ \begin{array}{cc} 1 & 1 \\ ~ & c\end{array}; {-z}\right] + \frac{1}{(d-1)} \ _{2}{F}_{1}\left[ \begin{array}{cc} 1 & 1 \\ ~ & d\end{array}; {-z}\right] \right\}   \,, 
\end{align}
with
\be
z = \frac{Q^2}{4 m_\pi^2}\,.
\ee

\section{Quark-connected Taylor coefficients}
\label{sec:TCs}

Here we tabulate the values of the Taylor coefficients employed in our analysis.  The light-quark connected $\Pi_j$s in Table~\ref{tab:LQPis} include corrections for finite-volume and discretization effects as described in Ref.~\cite{Chakraborty:2016mwy}.  The charm- and bottom-quark connected $\Pi_j$s in Table~\ref{tab:Piscb} have already been extrapolated to the continuum in Refs.~\cite{Donald:2012ga,Colquhoun:2014ica}.

\begin{table*}
    \caption{Light-quark-connected Taylor coefficients $\Pi^{(ud)}_j$ and strange-quark-connected Taylor coefficients $\Pi^{(s)}_j$ in units of $1/\mathrm{GeV}^{2j}$~\cite{Chakraborty:2014mwa,Chakraborty:2016mwy}. The quoted errors include statistics, the uncertainty on the vector-current renormalization factor,  the (correlated) uncertainty from setting the lattice spacing, and the uncertainty on the corrections.   The factor of the quarks' electromagnetic charges $(Q_u^2 + Q_d^2)$ is included in the definition of the $\Pi_j$s.  \vspace{1mm}}
    \label{tab:LQPis}
\begin{ruledtabular}
\begin{tabular}{ccccccccc}      
$\approx a$ (fm) & $\Pi^{(ud)}_1$ & $\Pi^{(ud)}_2$ & $\Pi^{(ud)}_3$ & $\Pi^{(ud)}_4$ & $\Pi^{(s)}_1$ & $\Pi^{(s)}_2$ & $\Pi^{(s)}_3$ & $\Pi^{(s)}_4$\\
\hline
0.15 & 0.0889(12)\textcolor{white}{a} & -0.1983(93) & 0.728(69) & -4.05(55) & 0.007387(83) & -0.00581(12) & 0.00509(17) & -0.00453(20)\\
0.12 & 0.08704(97) & -0.1884(80) & 0.682(62) & -3.82(49) & 0.007361(82) & -0.00584(12) & 0.00522(17) & -0.00477(21)\\
\end{tabular}
\end{ruledtabular}
\end{table*}

\begin{table}
    \caption{Charm- and bottom-quark-connected Taylor coefficients $\Pi^{(f)}_j$ in units of $1/\mathrm{GeV}^{2j}$~\cite{Donald:2012ga,Colquhoun:2014ica}.  The quoted errors include statistics and all systematics. The factors of the quarks' electromagnetic charges $Q_f^2$ are included in the definition of the $\Pi_j$s. \vspace{1mm}}
    \label{tab:Piscb}
\begin{ruledtabular}
\begin{tabular}{ccccc}
flavor & $10^3\, \Pi^{(q)}_1$ & $10^3\, \Pi^{(q)}_2$ & $10^3\, \Pi^{(q)}_3$ & $10^3\ \Pi^{(q)}_4$ \\
\hline
c & 1.840(49) & -0.1240(43) & 0.01081(43) & -1.030(41)e-3 \\
b & 0.0342(48) & -2.28(37)e-4 & 1.82(41)e-6 & -1.57(49)e-8 \\
\end{tabular}
\end{ruledtabular}
\end{table}

\newpage
\bibliographystyle{apsrev4-1} 
\bibliography{bibliography}

\end{document}